\documentclass[a4paper,11pt]{article}
\usepackage{pos}
\usepackage{tikz}
\usetikzlibrary{decorations.markings}
\usetikzlibrary{decorations.pathmorphing}
\definecolor{orangewolfram}{RGB}{224.584, 155.815, 36.223}
\definecolor{bluewolfram}{RGB}{93.9463, 129.229, 180.998}
\definecolor{greenwolfram}{RGB}{142.846, 176.35, 49.6957}

\def\cW{{\cal W}}
\def\da{{\dot{a}}}
\def\db{{\dot{b}}}
\def\dc{{\dot{c}}}
\def\dd{{\dot{d}}}
\def\cW{{\cal W}}
\def\cM{{\cal M}}
\def\be{\begin{equation}}
	\def\ee{\end{equation}}

\def\Pf{\mathrm{Pf}}

\numberwithin{equation}{section}

\title{The supergravity dual of a finite duality cascade}

\author*[a,b]{Pietro Moroni}
\author[a,b]{Fabrizo Aramini}
\author[c]{Riccardo Argurio}
\author[a,b]{Matteo Bertolini}
\author[a,b]{Eduardo Garc\'{\i}a-Valdecasas}

\affiliation[a]{SISSA,\\
Via Bonomea 265; I 34136 Trieste, Italy}
\affiliation[b]{INFN - Sezione di Trieste,\\
Via Valerio 2, 34127 Trieste, Italy}
\affiliation[c]{Physique Th\'eorique et Math\'ematique and International Solvay Institutes Universit\'e Libre de Bruxelles,\\
C.P. 231, 1050 Brussels, Belgium}

\emailAdd{pmoroni@sissa.it}
\emailAdd{faramini@sissa.it}
\emailAdd{Riccardo.Argurio@ulb.be}
\emailAdd{bertmat@sissa.it}
\emailAdd{egarciav@sissa.it}

\abstract{Cascading RG flows are characteristic of $\mathcal{N}=1$ gauge theories realized by D3-branes probing singularities in the presence of fractional branes. A celebrated example is the Klebanov-Strassler model, which exhibits an infinite cascade that ends with  confinement. In this work, we explore a related setup where the addition of an orientifold plane modifies the cascade structure: the RG flow now consists of a finite number of steps, originating from a UV fixed point with a finite number of degrees of freedom. We provide a supergravity solution dual to this flow, that reproduces all its salient features.}

\FullConference{Proceedings of the Corfu Summer Institute 2025 "School and Workshops on Elementary Particle Physics and Gravity" (CORFU2025)\\
27 April - 28 September, 2025\\
Corfu, Greece\\}

\begin{document}
\maketitle

\section{Introduction and Motivation}
\noindent
Gauge/gravity duality provides a powerful framework to study strongly coupled gauge theories via classical gravitational backgrounds. While its original formulation \cite{Maldacena:1997re,Gubser:1998bc,Witten:1998qj} relates conformal field theories to AdS geometries, many physically interesting systems exhibit non-trivial renormalization group (RG) flows and confinement. Extending holography to such settings remains an important direction.

A prominent example is the Klebanov--Strassler (KS) theory \cite{Klebanov:2000hb}, where fractional branes at the singularity of the conifold generate a cascading RG flow. The cascade proceeds through repeated Seiberg dualities and leads to confinement in the infrared (IR). However, the cascade continues indefinitely toward the ultraviolet (UV), reflecting the absence of a UV completion with a finite number of degrees of freedom.

In this work, we revisit this paradigm by introducing an orientifold plane into the conifold setup. This modification leads to a qualitatively different RG flow: the duality cascade becomes \emph{finite}. The theory interpolates between a UV regime described by a conformal field theory and an IR confining phase, providing a controlled example of a holographic flow with both a UV completion and confinement.

The underlying mechanism is that, in the presence of the orientifold, the effective number of fractional branes is not conserved along the flow. As a result, the difference between the ranks of the gauge groups decreases toward the UV and eventually vanishes, terminating the cascade after a finite number of steps. In the IR, the cascade ends in a confining vacuum, while in the UV the theory approaches a conformal manifold.

On the gravity side, we construct a supergravity solution that describes the flow. The geometry interpolates between a smooth geometry in the IR and an asymptotically $\text{AdS}_5 \times T^{1,1}$ background in the UV. Compared to the KS solution, the presence of the orientifold induces a non-trivial radial profile for the dilaton and modifies the behavior of the fluxes. These features are essential for reproducing the finite cascade structure.

We perform several non-trivial checks of the duality. In particular, the sequence of Seiberg dualities is matched with large gauge transformations of the $B_2$ field via Page charges, reproducing the expected evolution of gauge group ranks. The running of the gauge couplings, the pattern of $R$-symmetry breaking, and the UV central charge are all consistent between the field theory and gravitational descriptions.

The main results of this work can be summarized as follows:
\begin{itemize}
\item We identify a $\mathcal{N}=1$ gauge theory exhibiting a \emph{finite duality cascade}.
\item The theory admits a \emph{UV completion} in terms of a strongly coupled  superconformal field theory.
\item The IR dynamics is \emph{confining}, with a geometric realization in the dual background.
\item A supergravity solution captures (almost) the full RG flow and satisfies several quantitative checks of the duality.
\end{itemize}

This provides a controlled example of a holographic RG flow connecting a UV fixed point to a confining IR phase without requiring an infinite cascade.
\section{The Klebanov-Strassler model}
\label{sec:KS}
\noindent
We will start by introducing the KS theory and its holographic realization \cite{Klebanov:2000hb}, in order to then highlight the differences with our model. 

The KS theory is an $\mathcal{N} = 1$ $SU(N+M) \times SU(N)$ gauge theory, with four chiral superfields $A_1,\ A_2, \ B_1$ and $B_2$ in the bifundamental representation of the gauge group and a quartic superpotential, which can be written schematically as
\begin{equation}\label{eq:W}
    \mathcal{W} = h  \ \mathrm{Tr} \det_{r,u} (A_r B_u) \ .
\end{equation}
There is an $SU(2) \times SU(2) \times U(1)$ flavor symmetry, as well as a $U(1)_R$ R-symmetry.

The RG flow of this theory is rather complicated \cite{Strassler:2005qs}: the imbalance in the ranks of the two gauge groups makes the $\beta$ functions for the gauge couplings have opposite signs, which results in one of the gauge groups becoming strongly coupled at lower energies. We can then better describe the theory by means of a Seiberg duality on the strongly coupled gauge group:
\begin{equation}
    SU(N+M) \times SU(N) \rightarrow SU(N-M) \times SU(N) \ .
\end{equation}
The gauge singlets introduced by the Seiberg duality are massive due to the superpotential, and after integrating them out we are left with a quartic superpotential identical to \eqref{eq:W}, but with the dual quarks instead of the original ones. Therefore, the theory is self-similar after the Seiberg duality. Now the $SU(N)$ factor will be the strongly coupled gauge group, and we can do the same as before. The result is a cascade of Seiberg dualities. 

For $N = kM$, in the IR we are left with gauge group $SU(2M) \times SU(M)$: the moduli space of this theory admits a baryonic branch, where it reduces to $\mathcal{N} = 1$ $SU(M)$ SYM (which gives confinement and gaugino condensation) and a decoupled massless chiral multiplet (the Goldstone of the spontaneously broken baryonic symmetry). 

Instead, in the UV the cascade never stops: the theory is strongly coupled at arbitrarily high energy scales, and the number of colors goes to infinity with the energy. The UV completion is then given by a theory with an infinite number of degrees of freedom, as mentioned in the introduction.

The KS model is engineered as the low energy effective theory on a stack of $N$ regular D3 branes at the tip of the conifold in type IIB string theory, in the presence of $M$ fractional D3 branes. The conifold is a non-compact Calabi-Yau threefold, defined by the following equation in $\mathbb{C}^4$:
\begin{equation}
    z_1^2+z_2^2+z_3^2+z_4^2=0 \ .
\end{equation}
As explained in \cite{Candelas:1989js}, it is a cone over $T^{1,1}$, a five-dimensional space that is topologically $S^2 \times S^3$, and is usually parametrized by coordinates $r \geq 0$, $0 \leq \theta_i\leq \pi$, $0 \leq \varphi_i\leq 2\pi$, $0 \leq \psi\leq 4 \pi $.

For $M=0$, we find the Klebanov-Witten superconformal field theory \cite{Klebanov:1998hh}, which possesses a two-dimensional (complex) conformal manifold. This theory at large $N$ is holographically dual to type IIB supergravity on $AdS_5 \times T^{1,1}$.

The holographic dual for the $M \neq 0$ theory was found by Klebanov and Strassler in \cite{Klebanov:2000hb}. The singular conifold gets deformed so that the $S^3$ stays finite at $r=0$, and this smoothing of the geometry is the holographic dual of confinement. The symmetries of the field theory are reproduced by the isometries of the background, while the duality cascade is realized holographically as the running of the $F_5$ flux, which now depends on $r$. The difference in the ranks of the gauge groups $M$ is given by the constant $F_3$ flux and the logarithmic running of the gauge couplings is matched by the $B_2$ flux. 

However, the dilaton is constant, and this implies that the cascade is infinite in the UV. In the following, we will modify the KS construction to get a model with a finite duality cascade and find its gravity dual. 

\section{The orientifold field theory}
\label{sec:field_theory}
\noindent
Our field theory is obtained as an orientifold of the KS model implemented by an O7 plane embedded via the equation $z_4=0$ (this is the Kuperstein embedding, studied in \cite{Kuperstein:2004hy} in the case of D7 branes): the gauge group becomes $USp(N+M-2) \times USp(N)$, where $N$ and $M$ are always even. There are only two chiral multiplets $X_1$ and $X_2$, and we also have a quartic superpotential
\begin{equation}
\mathcal{W} = h J_{ab}J_{cd} J^{\db\dc}J^{\dd\da}\left[ (X_1)^a_\da (X_1)^b_\db (X_2)^c_\dc (X_2)^d_\dd - (X_1)^a_\da (X_2)^b_\db (X_2)^c_\dc (X_1)^d_\dd \right] \ ,
\end{equation}
where the indices $a$, $\da$ are the gauge group indices and $J$ is the symplectic invariant tensor.

The $SU(2) \times SU(2) \times U(1)_B$ global symmetry of the KS model is then broken to $SU(2)_F$. The classical $U(1)_R$ symmetry is instead broken by the ABJ anomaly to 
\begin{equation}
U(1)_R \, \rightarrow \, 
    \mathbb{Z}_{\text{gcd}(M,4)} \ ,
\end{equation}
which is $\mathbb{Z}_4$ for $M=0 \mod{4}$ and $\mathbb{Z}_2$ for $M=2 \mod{4}$.

The RG flow of the theory can again be described via a cascade of Seiberg dualities, just as in the KS case. However, in the case of symplectic gauge groups, Seiberg duality is modified \cite{Intriligator:1995ne}: the result is that the theory is self-similar after a cascade step with the different identification
\begin{equation}\label{eq: cascade_shifts}\begin{split}
	&N^\prime = N-M-2 \\
	&M^\prime = M+4 \ ,
\end{split}\end{equation}
with the roles of the two gauge groups now inverted. The cascade then takes the form
\begin{eqnarray*}
\label{eq: cascade steps}
&\cdots  \\
&\downarrow  \\ 
& USp(N+M-2)_1 \times USp(N+2M-8)_2  & \Delta (\text{rank}) = M-6  \\ 
&\downarrow  \\  &USp(N+M-2)_1 \times USp(N)_2 & \Delta (\text{rank}) = M-2  \\ 
&\downarrow  \\ 
& USp(N-M-2)_1 \times USp(N)_2 & \Delta (\text{rank}) = M+2  \\ 
&\downarrow  \\ 
&\dots 
\end{eqnarray*}

Differently from the KS cascade, we see that the difference between the ranks of the gauge groups at each duality step, $\Delta(\text{rank})\equiv |N^{(1)}-N^{(2)}|$, increases toward the IR and decreases towards the UV. This means that at high energies $M$ will go to zero, and the cascade will be finite.

Studying the $\beta$ functions of the theory, it can be shown that for $M=4$ and $M=2$, we actually have a conformal manifold of fixed points (of complex dimension one), similar to the Klebanov-Witten conformal manifold. 
Therefore, our theory admits as a UV completion any of the fixed points of the conformal manifold, and a relevant deformation of one of these fixed points triggers the cascade.

Instead, in the IR the rank difference gets bigger and bigger until a Seiberg duality is no longer possible. As for $USp$ gauge groups there are no baryons, one could naively expect that the moduli space would be entirely composed of mesonic branches, where the $SU(2)$ flavor symmetry would be broken.
However, we will see that an isolated vacuum with unbroken $SU(2)$ exists, so that we expect a purely geometric supergravity dual.

Let us consider our theory at the bottom of the cascade with gauge group $USp(N+M-2)\times USp(N)$. At the effective level, we can consider gauge invariants of the most strongly coupled gauge group, which we take to be the one of larger rank. These are
\begin{equation}
(M_{ij})_{\da\db} =J_{ab} (X_i)^a_\da (X_j)^b_\db\ .
\end{equation}
We can build a $2N\times 2N$ matrix as
\begin{equation}
\cM = \begin{pmatrix}
    M_{11} & M_{12} \\ M_{21} & M_{22}
\end{pmatrix}\ ,
\end{equation}
which has the property that $\cM^T=-\cM$, implying $M_{11}^T=-M_{11}$, $M_{22}^T=-M_{22}$ and $M_{12}^T=-M_{21}$.
The interesting case is $N=M+2$. In this case, the quantum corrected superpotential reads
\begin{equation}
\cW= h \ \mathrm{Tr} (M_{11} J M_{22} J - M_{12} J M_{21} J) + c \frac{\Pf \cM}{\Lambda^{2N-3}}\ ,
\end{equation}
where $\Pf$ is the Pfaffian.

The F-term equations for this superpotential admit the solution $\Pf \cM = 0$, which implies $M_{ij} = 0$ for each $i,j$. This vacuum is isolated from the others, as the other solutions have $\Pf \cM$ a nonzero constant. 
In this vacuum all mesons are massive and can be integrated out, so that we are left with pure $USp(M+2)$ SYM, which confines. This is the only vacuum whose dual background is expected to be purely geometrical, similarly to the baryonic branch of the KS theory. The supergravity solution we will find in the next section will describe precisely this vacuum.

Finally, one can show, using the repeated Seiberg dualities, that if we call the ranks of the two gauge groups at the bottom of the cascade $N^{(1)}_0=N_0+M_0-2$ and $N^{(2)}_0=N_0$, then at the nth step we will have
\begin{equation}\label{eq: shift at step k}\begin{split}
M_{n}&=M_0-4n  \\ 
N_{n}&=N_0+n\,M_0-2n^2\ . 
\end{split}\end{equation}
These expressions will be important because they are exactly reproduced by the dual supergravity background.

\section{Supergravity solution}
\noindent
The $USp(N+M-2) \times USp(N)$ gauge theory we have been discussing is obtained as the low energy theory on a stack of $N$ D3 branes at the tip of the conifold, in the presence of $M$ fractional D3 branes and an O7 plane. Without the O7, the setup reduces to the KS one.
The strategy is then to look for a deformation of the KS background that keeps into account the coupling of the O7 plane to the dilaton and the RR potential $C_0$. Since an O7 plane couples to these fields as a D7 brane of negative charge, the ansatz is very closely related to that of \cite{Benini:2007gx}, where the addition of fundamental flavors to the KS background via D7 branes was studied.

The ansatz for the metric in the Einstein frame is
\begin{equation}
    \label{eq: metric Ansatz}
    \begin{split}
        ds^2=&h(r)^{-\frac{1}{2}}dx_{1,3}^2+ h(r)^\frac{1}{2}\bigg[dr^2+e^{2G_1(r)} (\sigma_1^2+\sigma_2^2)+e^{2G_2(r)}((\omega_1+g(r)\sigma_1)^2+\\&(\omega_2+g(r)\sigma_2)^2) 
        +\frac{e^{2G_3(r)}}{9}(\omega_3+\sigma_3)^2
        \bigg] \, ,
    \end{split}
\end{equation}
where $dx_{0,3}^2$ is the Minkowski metric, the other coordinates are the same as for the (deformed) conifold and we have defined the following set of 1-forms
\begin{align}
\label{sigmaomega}
    &\sigma_1=d\theta_1 \quad , \quad  \sigma_2=\sin{\theta_1}d\varphi_1 \quad ,  \quad \sigma_3=\cos{\theta_1}d\varphi_1 \quad , \quad \omega_1=\sin{\psi}\sin{\theta_2}d\varphi_2+\cos{\psi}d\theta_2 \nonumber\\
&\omega_2=-\cos{\psi}\sin{\theta_2}d\varphi_2+\sin{\psi}d\theta_2 \quad , \quad \omega_3=d\psi+\cos{\theta_2}d\varphi_2 \ .
\end{align}
From now on, we set $\alpha'=g_s=1$. Introducing a new radial coordinate $\tau$, defined by $d\tau=3e^{-G_3}dr$, and solving the equations of motion, we get
\begin{equation}
    \label{eq: solution deformed conifold}
    \begin{split}
     &  h(\tau)=-\frac{\pi M_0^{\, 2}}{4\mu^\frac{8}{3}}
    \int\limits_{\tau}^{\infty} dx  \, H(x) \quad , \quad g^2 = 1 - e^{2(G_1-G_2)} \quad , \quad 
    e^{2G_1(\tau)}= \frac{1}{4} \mu^\frac{4}{3} \tilde{\Lambda}(\tau) \frac{\sinh^2\tau}{\cosh\tau} \ , 
    \\
    & e^{2G_2(\tau)} = \frac{1}{4} \mu^\frac{4}{3} \tilde{\Lambda}(\tau) \cosh\tau \quad , \quad 
        e^{2G_3(\tau)}= 6 \mu^{\frac{4}{3}} \frac{\tau-\tau_0}{\tilde{\Lambda}^2(\tau)}  \ ,
    \end{split} 
\end{equation}
where $\tau_0, \,\mu$ and $M_0$ are integration constants, $\tau>0$ (this actually follows from solving the equations of motion of the RR fields) and 
\begin{equation}
\label{eq: HtildeL}
    \begin{split}
       &H(x) = \frac{x\coth x-1}{(x-\tau_0)^2\sinh^2 x}\times 
    \frac{[-\cosh{(2x)}+4 x(x-\tau_0)+1-(x-2\tau_0)\sinh{(2x)}]}{\left[2\left(\sinh{(2x)}-x\right)(x-\tau_0)-\cosh{(2x)}+2x\tau_0+1\right]^\frac{2}{3}} \ , \\
       & \tilde{\Lambda}(\tau)=\frac{[2(\sinh{(
    2\tau)}-\tau)(\tau-\tau_0)-\cosh{(2\tau)}+2\tau\tau_0+1]^\frac{1}{3}}{\sinh{\tau}} \ .
    \end{split}
\end{equation}
Notice that the condition $g^2 = 1 - e^{2(G_1-G_2)}$ makes the metric symmetric under the $\mathbb{Z}_2$ generated by
\begin{equation}
    \theta_1 \longleftrightarrow \theta_2 \ , \quad \varphi_1 \longleftrightarrow \varphi_2 \ ,
\end{equation}
which is precisely the spacetime isometry gauged by the orientifold. In the following, everything will be written in the covering space, with the understanding that the physical space is actually its $\mathbb{Z}_2$ quotient.

The presence of the orientifold makes it so that the dilaton cannot be kept constant, and solving its equation of motion we find
\begin{equation}\label{eq: dilaton solution}
    e^\phi=\frac{\pi}{2}\frac{1}{\tau-\tau_0} \ . 
\end{equation}
For this to make sense, we need to take $\tau_0 <0$, which means that the point at which the dilaton diverges will not be part of the geometry, since $\tau$ needs to be positive.
Moreover, the O7 spoils the Bianchi identity for $F_1$ by acting as a localized source for $C_0$. Finding a solution for a localized source is very hard, therefore we simplify the problem by smearing the orientifold plane, that is we impose
\begin{equation}
    dF_1= -\frac{2}{\pi}(\omega_1\wedge\omega_2-\sigma_1\wedge\sigma_2) \ , 
\end{equation}
which implies
\begin{equation} \label{eq: F_1}
    F_1=-\frac{2}{\pi}(\omega_3+\sigma_3) \ . 
\end{equation}

The ansatz for the NSNS and RR 3-form field strengths, as well as the one for the RR 5-form field strength, are rather complicated, so we will not write them down here. The interested reader may find them in \cite{Aramini:2025twg}. 
What we will actually need to match supergravity and gauge theory are the fluxes of the RR forms, which correspond to the effective number of regular and fractional D3 branes. We will see that they correspond under the gauge/gravity dictionary to the ranks of the gauge groups. Their expression is
\begin{equation}\label{eq:NM}\begin{split}
   & N_{eff}(\tau) = \frac{1}{(4 \pi^2)^2} \int_{T^{1,1}} 
    F_{5} = N_0+
    \frac{M_0^{\, 2}}{\pi} \biggl( f - (f - k) F - \frac{2}{\pi} f k \biggr) \ , \\
    & M_{eff}(\tau) = \frac{1}{4 \pi^2} \int_{S^{3}} F_{3} = M_0 \biggl(1 - \frac{2}{\pi} (f + k) \biggr) \ ,
\end{split}\end{equation}
where the equations of motion for the RR fields fix 
\begin{equation}\label{eq: solution of fluxes equation}
    \begin{split}
        e^{-\phi} f &= \frac{\tau\coth{\tau} - 1}{2 \sinh {\tau}}(\cosh{\tau}- 1)\ ,\\
        e^{-\phi} k &= \frac{\tau\coth{\tau} - 1}{2 \sinh {\tau}}(\cosh{\tau}+ 1)\ , \ ,\\
        F &= \frac{\sinh{\tau} - \tau}{2 \sinh{\tau}} \ ,
    \end{split}
\end{equation}
and $N_0$ is another integration constant. As we can see, differently from the KS case, also the 3-form flux varies with $\tau$. This is the SUGRA realization of the fact that on the field theory side the difference in the ranks of the two gauge groups changes at each duality step.

Let us now study our solution in the regimes of greatest interest, that is for small and large $\tau$, which correspond on the field theory side to the IR and UV, respectively. First of all, for $\tau \sim 0$, we have
\begin{equation}
     e^{\phi} \rightarrow - \frac{\pi}{2}\frac{1}{\tau_0}, \qquad e^{-\phi} f \sim \frac{1}{12}\tau^3, \qquad e^{-\phi} k \sim \frac{1}{3}\tau, \qquad F \sim \frac{1}{12}\tau^{2} \ ,
\end{equation}
which imply, using \eqref{eq:NM}, that
\begin{equation}
    M_{eff} (\tau) \underset{\tau \sim 0}{\to} M_0, \qquad N_{eff} (\tau) \underset{\tau \sim 0}{\to} N_0  \ .
\end{equation}
Therefore, $N_0$ and $M_0$ have to be interpreted as the number of regular and fractional branes in the deep IR, where the gauge theory confines. In the previous section, we have seen that in the vacuum we are interested in, the theory reduces to $USp$ SYM, which has a single gauge group. This suggests that we should set $N_0 = 0$, so that only fractional branes are left at $\tau = 0$ and supergravity can match the gauge theory dynamics. The warp factor goes to a constant as $\tau$ goes to zero, and the six-dimensional metric reduces to
\begin{equation}
    ds^2_6 \underset{\tau \sim 0}{\sim} \left(\frac{\mu^4}{3}|\tau_0|\right)^\frac{1}{3}\left[\frac{1}{2}d\tau^2+d\Omega_3^2+\frac{1}{4}\tau^2d\Omega_2^2\right] \ ,
\end{equation}
where $d\Omega_3^2$ and $d\Omega_2^2$ are the metric on the $S^3$ and $S^2$, respectively. We can see that the $S^2$ shrinks to zero size as $\tau \rightarrow 0$, while the $S^3$ remains finite, just as in the KS solution. As mentioned in section \ref{sec:KS}, the regularity of the geometry at $\tau = 0$ is what gives the holographic realization of confinement.

Actually, even though the metric looks perfectly smooth at $\tau = 0$, there is a curvature singularity that is absent in the KS solution: the Ricci scalar diverges as $\frac{1}{\tau}$. We interpret this singularity as a consequence of the smearing approximation. This singularity is, however, a "mild" singularity, in the sense that it does not spoil any simple holographic computation related to confinement (for example the area law of the holographic Wilson loop).

At large $\tau$, we would expect our geometry to approach the undeformed conifold, since the dual gauge theory hits a fixed point in the extreme UV. Expanding \eqref{eq: solution of fluxes equation}, one gets 
\begin{equation}
\label{UV functions}
    e^{\phi} \sim  \frac{\pi}{2\tau},\qquad e^{-\phi} f \sim e^{-\phi} k \sim \frac{\tau}{2}, \qquad F \sim \frac{1}{2} \ ,
\end{equation}
so that
\begin{equation} \label{eq: Meff and Neff in the UV}
    M_{eff} (\tau)  \xrightarrow[\tau \rightarrow \infty]{}  0, \quad
    N_{eff} (\tau) \xrightarrow[\tau \rightarrow \infty]{} \frac{M_0^{\, 2}}{8} \ .\end{equation}
    
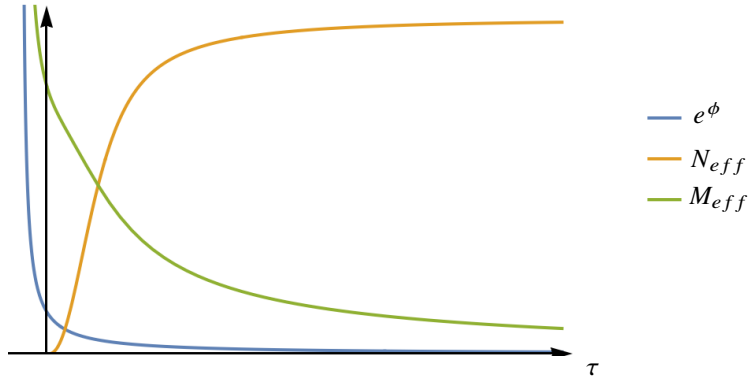
\begin{figure}[h!] \label{fig:1}
  \centering
  \begin{tikzpicture}[scale=0.9, transform shape]
    \node[anchor=south west, inner sep=0] (full) at (9,0)
      {\includegraphics[width=0.55\textwidth]{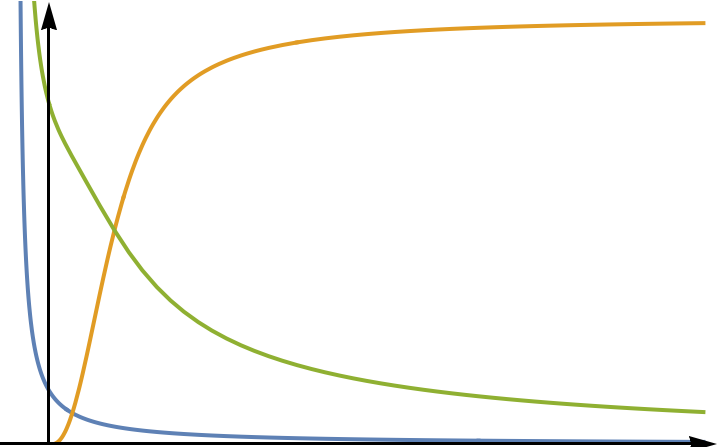}};

    \draw [line width = 1.2pt, bluewolfram] (18.4,3.5) -- (18.9,3.5);
    \node at (19.28,3.58) {$e^{\phi}$};
    \draw [line width = 1.2pt, orangewolfram] (18.4,2.9) -- (18.9,2.9);
    \node at (19.45,2.85) {$N_{eff}$};
    \draw [line width = 1.2pt, greenwolfram] (18.4,2.3) -- (18.9,2.3);
    \node at (19.45,2.3) {$M_{eff}$};
    \node at (17.6,-0.2) {$\tau$};
        
  \end{tikzpicture}
  \caption{$e^{\phi}$, $M_{eff}$ and $N_{eff}$ as functions of $\tau$}
  \label{fig: figura_1}
\end{figure}

Again, this matches the field theory expectation that the difference in the ranks of the gauge groups gets smaller and smaller towards the UV. We will see the precise relation between RR fluxes and gauge theory ranks shortly. In figure \ref{fig: figura_1} we show the behavior of $e^{\phi}$, $M_{eff}$ and $N_{eff}$ as a function of $\tau$: we see that, as expected, $M_{eff}$ is an increasing function of $\tau$, while $N_{eff}$ is decreasing. 

The asymptotic expansion for the six-dimensional metric is easier to do using the $r$ coordinate instead of $\tau$ (the relation between the two is monotonic), and it reads
\begin{equation}
\label{UVmetric}
    ds^2_6\underset{r\to\infty}{\sim} [dr^2+\frac{r^2}{6}(\sigma_1^2+\sigma_2^2+\omega_1^2+\omega_2^2)+\frac{r^2}{9}(\sigma_3+\omega_3)^2] \ ,
\end{equation}
which is the metric of the undeformed conifold. The warp factor is
\begin{equation}
\label{eq: UV asymp warp factor}
     h(r)\underset{r\to\infty}{\sim} \frac{27\pi}{4r^4} \frac{M_0^{\, 2}}{8} \left[1-\frac{1}{2\log{r}}\right] \ ,
\end{equation}
which, up to  logarithmic-suppressed  corrections, gives the metric of $AdS_5 \times S^5$. This shows that the supergravity solution asymptotes to pure $AdS_5 \times T^{1,1}$ geometry  with $\frac{M_0^2}{8}$ units of $F_5$ flux, in agreement with the existence of a UV fixed point in the dual field theory. 

As $r$ goes to $\infty$, however, $e^{\phi} N_{eff}$ goes to $0$, which means that the curvature in the string frame diverges. Therefore, if we were to interpret the UV limit of our solution as some holographic superconformal field theory, it would be outside the regime in which classical supergravity is a good approximation. However, one can show that that for every choice of $M_0$ we can choose the integration constant $\tau_0$ such that the supergravity approximation remains valid up to parametrically large $r$. In particular, it can be a good description of the dual field theory at energies much larger than the scale at which the duality cascade stops.
\section{Gauge/gravity duality checks}
\label{sec: checks}
\begin{figure}[h!]
  \centering
  \begin{tikzpicture}[scale=1.15, transform shape]
    \node[anchor=south west, inner sep=0] (full) at (9,0)
      {\includegraphics[width=0.55\textwidth]{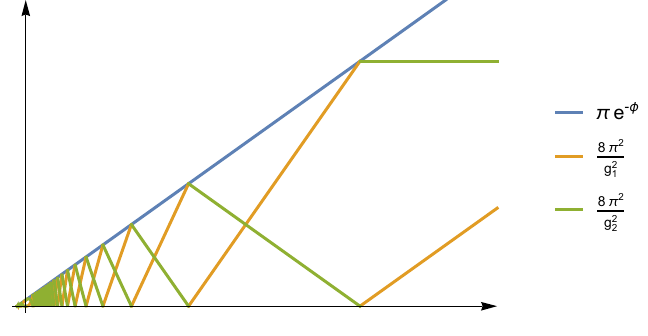}};

    \node at (15.6,0.15) {$\tau$};
        
  \end{tikzpicture}
  \caption{Supergravity prediction for the inverse squared gauge couplings. We see the appearance of the duality wall at $\tau = \tau_0$. For small enough values of $\tau$ the plot cannot be trusted, as equation \eqref{eq:couplings} is not valid anymore.}
  \label{fig: figura_2}
\end{figure}

\noindent
The starting point for the matching of gauge theory and supergravity quantities is the dictionary 
\begin{equation}\label{eq:couplings}
    \frac{8 \pi^{2}}{g_{1}^{2}}  + \frac{8 \pi^{2}}{g_{2}^{2}} = \pi e^{- \phi}  \quad , \quad 
 \frac{8 \pi^{2}}{g_{1}^{2}} - \frac{8 \pi^{2}}{g_{2}^{2}}  = 2 \pi e^{- \phi} \biggl[ \left( \frac{1}{4 \pi^2} \int_{S^{2}} B_{2} \right) \bmod 1 - \frac{1}{2} \biggr]  \ .
\end{equation}
Strictly speaking, these relations are only valid in the UV, where we are close enough to the conformal manifold. In figure \ref{fig: figura_2}, we show a plot of the squared inverse gauge couplings as predicted by supergravity.

We see that for integer values of the $B_2$ flux, one of the two gauge couplings diverges (the inverse goes to zero). Moreover, the integral of $B_2$ is not a gauge invariant quantity, as it shifts by an integer under a large gauge transformation of $B_2$. Since the two configurations are gauge-equivalent, they must be physically equivalent, which leads to the identification of a large gauge transformation of $B_2$ with a Seiberg duality on the field theory side. Then, every time the flux of $B_2$ hits an integer as $\tau$ varies, the effective description changes, and the ranks of the $USp$ groups get shifted. Since this happens many times from $\tau = 0$ to $\tau = +\infty$, the result is a cascade of Seiberg dualities. 

Since $e^{-\phi} \rightarrow \infty$ for large $\tau$, the cascade steps will take longer and longer going towards the UV, until eventually the cascade will stop, perfectly matching the gauge theory expectation of a finite cascade. Finally, $e^{-\phi}$ goes to zero as $\tau$ goes to $\tau_0$, which means that $\tau_0$ is an accumulation point for the Seiberg dualities. However, this pathological behavior is actually not present in the solution, as $\tau_0 < 0$ and the geometry ends at $\tau=0$, effectively cloaking the singularity.

\subsection{Gauge theory ranks and $\beta$ functions}
\noindent
We now want to precisely match the ranks of the gauge groups to supergravity quantities. As we have now said multiple times, we know that the effective number of branes is related to the gauge theory ranks. However, the ranks must be integer numbers, while the RR fluxes are not quantized in our background, as they change continuously with $\tau$. Natural quantized objects in the theory are the Page charges
\begin{equation}
\label{Page chs}
    Q_{D5} = \frac{1}{4 \pi^{2}} \int_{S^{3}} dC_{2} \quad , \quad Q_{D3} =\frac{1}{(4\pi^2)^2}\int_{T^{1,1}} dC_{4} \ ,
\end{equation}
where 
\begin{equation}\label{eq:potentials}
    dC_{2} = F_{3} - B_2 \wedge F_1, \quad dC_{4} = F_{5} - B_2 \wedge F_3 + \frac{1}{2} B_2 \wedge B_2 \wedge F_1 \ .
\end{equation}
However, the Page charges are not gauge invariant. The natural prescription is to match the ranks of the gauge groups at the scales where $M_{eff}$ and $N_{eff}$ are equal to one of the corresponding Page charges, which one can easily check are exactly the values at which the flux of $B_2$ is an integer (that is, precisely where we have to perform a Seiberg duality). 
Then, letting $\tau_n$ be the value of the radial coordinate at which the flux of $B_2$ hits an integer for the $n^{th}$ time, we find that
\begin{equation}\begin{split}\label{eq:ranks}
    &M_{eff} (\tau_{n}) = Q_{D5}^n = M_0 - 4n = N^{(1)}_n-N^{(2)}_n+2 = M_n \\
    &N_{eff} (\tau_{n}) = Q_{D3}^n = nM_0 - 2n^2 = N^{(2)}_n = N_n \ ,
\end{split}\end{equation}
where $N^{(1)}_n$ and $N^{(2)}_n$ are the number of colors for the two gauge groups at the $n^{th}$ step of the cascade going toward the UV, which give the effective description for $\tau \in (\tau_n,\tau_{n+1})$. Eq.~\eqref{eq:ranks} precisely matches the ranks at each step and their shifts, given in \eqref{eq: shift at step k} (for $N_0=0$).

We can now go one step further and use \eqref{eq:couplings} to compute the field theory $\beta$ functions at high energies (where \eqref{eq:couplings} is valid). For large $\tau$, we have \cite{Aramini:2025twg}
\begin{equation}\label{eq:asymptotic_B}
    \frac{1}{4 \pi^2} \int_{S^{2}} B_{2} \underset{\tau \to \infty}{\sim}  \frac{M_0}{\pi} f(\tau) \ ,
\end{equation}
which gives the following expressions for the gauge couplings as functions of the radial coordinate: 
\begin{equation}
\begin{split}
\label{eq:couplings_mod}
\frac{8 \pi^{2}}{g_{1}^{2}} \underset{\tau \to \infty}{\sim} M_0 e^{-\phi} f - n \pi e^{-\phi} \quad , \quad
\frac{8 \pi^{2}}{g_{2}^{2}} \underset{\tau \to \infty}{\sim}  - M_0 e^{-\phi} f + (n+1) \pi e^{-\phi} \ ,
\end{split}
\end{equation}
where $n =  \left \lfloor {\frac{M_0}{\pi} f(\tau)} \right \rfloor$ is the step of the cascade at which $g_1$ and $g_2$ have to be interpreted as the couplings of the effective description valid for $\tau \in (\tau_n, \tau_{n+1})$. One can show that in the UV the energy-radius relation is given by 
\begin{equation}
    \log \mu = \frac{1}{3} \tau + \text{const} \ ,
\end{equation}
which implies
\begin{equation}\label{eq:beta_derivative}
    \frac{d}{d \log \mu} = 3 \frac{d}{d \tau} \ .
\end{equation}
Using \eqref{eq:beta_derivative}, we have the $\beta$ functions
\begin{equation}\begin{split}\label{eq:cascade_plot}
    &\beta_1 \sim 3 \frac{d}{d \tau} \frac{8 \pi^2}{g_1^2} \sim  \sim 3 \frac{d}{d \tau} \biggl( \frac{M_0 \tau}{2} - 2 n \tau  \biggr) \sim \frac{3}{2} (M_0-4n) = \frac{3}{2} M_n  \ , \\ &\beta_2 \sim 3 \frac{d}{d \tau} \frac{8 \pi^2}{g_1^2} \sim 3 \frac{d}{d \tau} \biggl(- \frac{M_0 \tau}{2} + 2 (n+1) \tau  \biggr) \sim -\frac{3}{2} (M_0-4n)+6 = - \frac{3}{2} (M_n - 4) \ ,
\end{split}\end{equation}
where we have used that for large $\tau$ $e^{-\phi} \sim \frac{2}{\pi} \tau$ and $e^{-\phi} f \sim \frac{\tau}{2}$, and the relations \eqref{eq:ranks} to express the result in terms of the number of colors of the $USp(N_n + M_n-2) \times USp(N_n)$ effective description. 
Equations \eqref{eq:cascade_plot} match the gauge theory result at leading order in $\frac{M}{N}$ \cite{Aramini:2025twg}. Then, in the UV, the $\beta$ functions are constant in each energy range, which means that for large $\tau$ the RG flow will be logarithmic (just like in the KS case).

\subsection{R-symmetry breaking pattern}
\noindent
We now want to see how the breaking of the field theory $U(1)$ R-symmetry via the ABJ anomaly is realized in supergravity. As mentioned in section \ref{sec:field_theory}, our $USp(N+M-2) \times USp(N)$ theory possesses a $U(1)$ R-symmetry, which is broken at the quantum level to the discrete subgroup $\mathbb{Z}_{\text{gcd}(M,4)}$. 

In the supergravity background, the R-symmetry is realized as the $U(1)$ isometry of the metric given by $\psi \rightarrow \psi+\alpha$, with $\alpha \in [0,4 \pi]$. Despite being a symmetry of the metric, this $U(1)$ is not a symmetry of the whole solution: the RR potentials $C_0$ and $C_2$ are not invariant. To be precise, since $d F_1 \neq 0$, to properly define $C_0$ we need to restrict to a submanifold $\mathcal{C}_4$ inside our background, defined by $\theta_1=\theta_2=\theta$, $\varphi_1=2\pi-\varphi_2=\varphi$, where $F_1$ is closed. On this submanifold, the RR potentials are (up to gauge transformations)
\begin{equation}\label{eq: effective RR potentials}
    C_0=-\frac{2}{\pi}\psi \quad, \quad C_2=\frac{M_0}{2} \psi\, \sin{\theta}\,d\theta\wedge d\varphi~.
\end{equation}
Being both linear in $\psi$, we can immediately see that they will not be invariant under $\psi \rightarrow \psi + \alpha$, for a generic value of $\alpha$. However, for specific values, the potentials will be mapped to a gauge equivalent configuration, resulting in a symmetry of the full background. The (large) gauge transformations of the potentials are given by
\begin{align}
\label{eq: large gauge C0}
    C_0&\to C_0+\lambda_0 ,  &\lambda_0=2k \ , \quad k \in \mathbb{Z} \ , \\
\label{eq: large gauge C2}
     C_2&\to C_2+\lambda_2,  &\lambda_2=2\pi\Omega_2 n \, , \quad n \in \mathbb{Z} \ ,
\end{align}
where $\Omega_2 = \frac{1}{2}(\sin{\theta_1}d\theta_1 d\varphi_1-\sin{\theta_2}d\theta_2 d\varphi_2)$ is the volume form on the 2-sphere of the background. The quantization of the gauge transformations has a factor of 2 with respect to the naive expectation due to the presence of the orientifold (the D5 and D7 charges are multiplied by 2 in the covering space). Then, we have a symmetry when
\begin{equation}
\begin{split}
    &\Delta C_0=-\frac{2}{\pi} \alpha = 2k \implies\alpha= \pi k, \quad k=0,1,2,3 \\
    &\Delta C_2= \frac{M_0}{2} \Omega_2 \ \alpha = 2 \pi \Omega_2 n \implies \alpha= \frac{4\pi}{M_0}n, \quad n=0,1,...,M_0-1 \ ,
\end{split}
\end{equation}
where we have used the fact that $\Omega_2 = \sin \theta \ d \theta \wedge d \varphi$ on $\mathcal{C}_4$ . \\
This implies the breaking
\begin{equation}
\label{R-sym breaking 1}
    U(1)_R\to \mathbb{Z}_{gcd(M_0,4)}~,
\end{equation}
in exact agreement with the field theory result (recall that $M$ changes by 4 at each step of the cascade, so that $\text{gcd}(M,4) = \text{gcd}(M_0,4)$).
\subsection{Holographic central charge}
\noindent
The final check we want to provide is the holographic computation of the $a$ and $c$ charges of the $\mathcal{N}=1$ superconformal field theories that give the UV completion of our model. In a ${\cal N}=1$ SCFT $a$ and $c$ are related to the R-charges, and an explicit computation for both the possible UV completions of our model ($M_{UV} = 2$ or $M_{UV} = 4$) gives, at leading order in $N_{UV}$,
\begin{equation}
\label{ac_qft}
    a=c=\frac{27}{128} N_{\text{UV}}^2 \ .
\end{equation}
While central charges can be rigorously defined only at conformal fixed points, in holography one can define a monotonic function $c(\tau)$, built from the components of the metric, which measures the number of effective degrees of freedom of the theory along the flow. Its explicit form in our background is given by
\begin{equation}
\label{holc0}
    c(\tau) = \frac{4}{3 \pi^2} \frac{e^{2 G_1(\tau)+2 G_2(\tau)+ 4 G_3(\tau)} h^2(\tau)}{\left(4 G_1'(\tau)+4 G_2'(\tau)+2 G_3'(\tau) +h'(\tau)/h(\tau)\right)^3} \ .
\end{equation}
This is a monotonically decreasing function as $\tau$ decreases, and it correctly approaches $c = \frac{27}{128} N_{\text{UV}}^2$ as $\tau \rightarrow \infty$. Moreover, it goes to zero as $\tau$ goes to zero, as expected since our theory is gapped, and therefore has no IR degrees of freedom. 

We can compare these results with those obtained in the KS solution: also in that case, the $c$-function \eqref{holc0} vanishes in the IR, reflecting the fact that the vacuum on the baryonic branch is gapped (except for the Goldstone chiral superfield  of the spontaneously broken baryonic symmetry, whose contribution is not captured in the supergravity limit). \\
However, in contrast to our setup, it  diverges quadratically in $\tau$ in the UV, consistently with the expected UV completion of the KS theory involving an infinite number of degrees of freedom.

\bibliographystyle{JHEP}
\bibliography{main}

\end{document}